\documentclass[twocolumn, showpacs, prb, superscriptaddress, citeautoscript]{revtex4}
\usepackage{graphicx}% Include figure files

\usepackage{dcolumn}% Align table columns on decimal point

\usepackage{color}
\usepackage{bm}% bold math

\definecolor{green}{rgb}{0.0,0.60,0.0}

\usepackage[normalem]{ulem}

\begin{document}

\title{Effective Mass and Spin Susceptibility of Dilute Two-Dimensional Holes in GaAs}

\date{\today}

\author{YenTing\ Chiu}
\affiliation{Department of Electrical Engineering, Princeton University, Princeton, NJ 08544}

\author{Medini\ Padmanabhan}
\affiliation{Department of Electrical Engineering, Princeton University, Princeton, NJ 08544}

\author{T.\ Gokmen}
\affiliation{Department of Electrical Engineering, Princeton University, Princeton, NJ 08544}

\author{J.\ Shabani}
\affiliation{Department of Electrical Engineering, Princeton University, Princeton, NJ 08544}

\author{E.\ Tutuc}
\affiliation{Department of Electrical Engineering, Princeton University, Princeton, NJ 08544}

\author{M.\ Shayegan}
\affiliation{Department of Electrical Engineering, Princeton University, Princeton, NJ 08544}

\author{R.\ Winkler}
\affiliation{Department of Physics, Northern Illinois University, DeKalb, Illinois 60115}
\affiliation{Materials Science Division, Argonne National
Laboratory, Argonne, Illinois 60439}

\begin{abstract}

We report effective hole mass ($m^{*}$) measurements through
analyzing the temperature dependence of Shubnikov-de Haas
oscillations in dilute (density $p \sim 7 \times 10^{10}$ cm$^{-2}$, $r_{s} \sim  6$)
two-dimensional (2D) hole systems confined to a 20 nm-wide, (311)A
GaAs quantum well. The holes in this system occupy two
nearly-degenerate spin subbands whose $m^{*}$ we measure to be $\sim $ 0.2 (in units of the free electron mass). Despite the relatively large $r_{s}$ in our 2D system, the measured $m^{*}$ is in reasonably good agreement with the results of our energy band calculations which do not take interactions into account. We then apply a sufficiently strong parallel
magnetic field to fully depopulate one of the spin subbands, and
measure $m^{*}$ for the populated subband. We find that this latter
$m^{*}$ is close to the $m^{*}$ we measure in the absence of
the parallel field. We also deduce the spin susceptibility of the 2D hole system from the depopulation field, and conclude that the susceptibility is enhanced by about 50$\%$ relative to the value expected from the band calculations.

\end{abstract}

\maketitle

\section{Introduction}

The ground state properties of a low disorder and dilute two-dimensional (2D) system of charged particles are dominated by the Coulomb interaction. For a 2D particle density $p$, the interaction strength is characterized by the parameter $r_{s}=1/\sqrt{{\pi}pa^{*}_{B}}$, the average inter-particle spacing measured in units of the effective Bohr radius ($a^{*}_{B}$), and increases as the system is made more dilute. In Fermi liquid theory, interacting particles can be treated as non-interacting quasi-particles with a re-normalized effective mass ($m^{*}$) and spin susceptibility, ${\chi}^* \propto g^{*} m^{*}$, where $g^*$ is the Lande g-factor. In the highly interacting, dilute regime ($r_{s} \gtrsim 3$), ${\chi}^*$ and $m^{*}$ are typically enhanced compared to the band values and increase with increasing $r_s$, as confirmed both theoretically \cite{CeperleyPRB1989, AttaccalitePRL2002, AsgariSSC2004, DePaloPRL2005, GangadharaiahPRL05, ZhangPRL05, ZhangPRB05, AsgariPRB2006} and experimentally \cite{SmithPRL72, OkamotoPRL1999, PanPRB99, ShashkinPRL2001, VitkalovPRL2001, PudalovPRL02, ShashkinPRB02, PrusPRB2003, ShashkinPRL03, TutucPRB2003, ZhuPRL2003, VakiliPRL04, TanPRL2005, TanPRB2006, GokmenPRB07}. Besides $r_{s}$, the spin and/or valley degrees of freedom also play an important role in the re-normalization of $m^{*}$ and ${\chi}^*$ since they modify the exchange interaction \cite{GangadharaiahPRL05, ZhangPRL05, ShashkinPRL03, PadmanabhanPRL2008, GokmenPRL2008, GokmenPRB2009, GokmenPRB2010}. In particular, it was recently demonstrated that when a 2D electron system is fully spin and valley polarized, $m^{*}$ is \textit{suppressed} compared to its band value \cite{PadmanabhanPRL2008, GokmenPRL2008, GokmenPRB2009, GokmenPRB2010}. This suppression was reproduced in subsequent, numerical theoretical work \cite{AsgariPRB2009, DrummondPRB2009}.

In this work we present measurements of $m^{*}$ in a dilute 2D \textit{hole} system (2DHS) confined to a (311)A GaAs quantum well. We determine $m^{*}$ through analysis of the temperature dependence of the Shubnikov-de Haas (SdH) oscillations. While $m^{*}$ in GaAs 2D holes has been measured via cyclotron resonance experiments \cite{StormerPRL1983, HirakawaPRB1993, ColeJPCM1997, ColePRB1997, PanAPL2003, LuAPL2008, RachorPRB2009}, there are no reports of $m^{*}$ measurements in dilute 2DHSs through transport experiments. The latter are useful because $m^{*}$ deduced from transport measurements should reflect interaction effects while, due to the Kohn theorem \cite{KohnPR1964}, $m^{*}$ measured in cyclotron resonance experiments does not.

The 2D holes in GaAs differ from their 2D electron counterparts in several notable aspects. First, the band value of $m^{*}$ for holes is much larger than for electrons. The larger $m^{*}$ results in a larger $r_{s}$ so that one would expect the 2DHS to be a more interacting system compared to a 2D electron system at the same density. Second, unlike electrons, the energy band structure of holes is nonparabolic. This is closely related to the fact that 2D holes have effective spin $j = 3/2$ rather than $j = 1/2$, which is the case for electrons. Third, the spin-orbit interaction resulting from the lack of inversion symmetry in GaAs typically leads to a more pronounced splitting of the energy bands (at finite wave vectors) in the 2DHS. In an in-plane magnetic field $B_\|$ this is complemented by Zeeman splitting that affects even the states at $k=0$. The resulting two spin subbands, which we refer to in this paper as $p^+$ and $p^-$ (see Fig.\ 1 right inset), in general have different energy surfaces, populations, and effective masses.

In our low-density 2DHS which is confined to a nearly symmetric GaAs quantum well, the spin-orbit interaction induced splitting is small and we cannot experimentally resolve it. From our measurements, we find that $m^{*}$ has a value of $\sim 0.2$ (in units of free electron mass, $m_e$), and slightly decreases as the density is decreased. Both the value of $m^{*}$ and its density dependence are in agreement with the results of energy band calculations which do not take interaction into account. This lack of $m^{*}$ enhancement relative to the results of the band calculations neglecting exchange-correlation is surprising. These results corroborate qualitatively with previous spin susceptibility measurements in 2DHSs \cite{WinklerPRB2005}, which were also found to be close to calculated values in the absence of electron-electron interaction \cite{Pinczuk86}. While a satisfactory theoretical explanation for these observations is currently lacking, it is apparent that the holes' band structure and their $j = 3/2$ effective spin significantly alter the impact of electron-electron interaction in comparison with spin $j = 1/2$ particles.

In our study, we also apply a large magnetic field ($B_{\|}$) parallel to the 2D plane to induce a polarization of the carriers' spin and thus to separate the two spin subbands. By applying a sufficiently large field, we are able to fully depopulate one of the spin subbands, and measure $m^{*}$ for the occupied subband. The measured $m^{*}$ is close to the value measured when $B_{\|} = 0$ and does not appear to be affected by the large $B_{\|}$ which should in principle couple to the holes' orbital motion and lead to an increase in $m^{*}$ \cite{TutucPRB2003}. Finally, from the value of $B_{\|}$ at which the minority spin subband is depopulated, we deduce a value for the 2D holes' spin susceptibility which is about 50$\%$ enhanced with respect to the band value.

\section{Sample Parameters and Experimental Details}

Our sample is grown on a GaAs (311)A substrate and consists of a 20 nm-wide GaAs quantum well modulation-doped with Si. As grown, the sample has density $p = 7.2 \times 10^{10}$ cm$^{-2}$ and mobilities of 35 and 55 m$^{2}$/Vs along the $[01\bar{1}]$ and $[\bar{2}33]$ directions, respectively \cite{Footnote1}. We performed measurements on a sample with van der Pauw geometry. A metallic front gate was deposited to control the carrier density, $p$, which we determined from the field positions of the resistance minima, the frequency of the SdH oscillations and from the Hall resistance. Values of $p$ in our sample are in the range from 3.7 to $9.0 \times 10^{10}$ cm$^{-2}$. We made longitudinal ($R_{xx}$) and Hall ($R_{xy}$) resistance measurements in a pumped $^{3}$He system with a base temperature $T = 0.3$ K. The sample was mounted on a single-axis tilting stage that could be rotated using a computer-controlled stepper motor, in order to change the angle ($\theta$) between the sample plane and the magnetic field. The measurements were done using low-frequency lock-in techniques.

\begin{figure}
\centering
\includegraphics[scale=0.85]{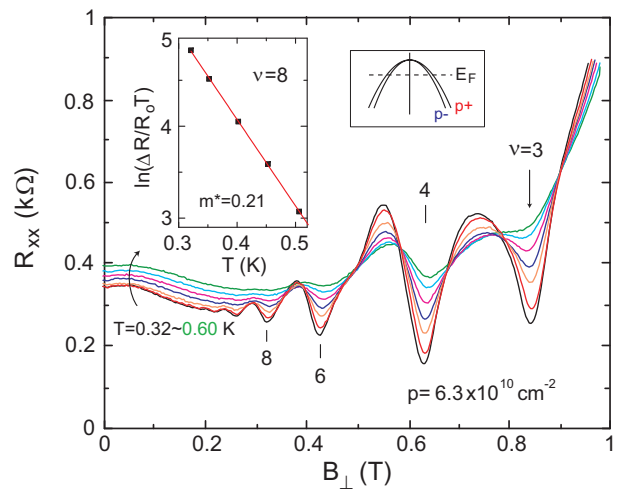}
\caption{Magnetoresistance traces at a density of
$6.30 \times 10^{10}$ cm$^{-2}$. The traces were taken at various temperatures
as shown in different colors. Left inset shows the Dingle fit at
$\nu = 8$ assuming a constant $\tau_q$ and $R_o$. Right inset is a
schematic picture showing the band structure of the GaAs hole system at zero magnetic field.}\end{figure}

To deduce $m^{*}$, we analyzed the $T$-dependence of the amplitude ($\Delta R$) of the SdH resistance
oscillations using the standard Dingle expression:\cite{DinglePRSL52} $\Delta
R/R_{0} \propto \exp (-\pi/\omega_{c}\tau_{q}) \,\xi/\sinh(\xi)$, where the
factor $\xi/\sinh(\xi)$ represents the \textit{T}-induced damping
($\xi=2\pi^{2}k_{B}T/\hbar\omega_{c}$), and
\textit{$\omega_{c}=eB_{\bot}/m^{*}$} is the cyclotron frequency;
$B_{\bot}$ is the perpendicular component of the magnetic field,
$R_{0}$ is the non-oscillatory component of the resistance near a
SdH oscillation, and $\tau_{q}$ is the single-particle (quantum)
lifetime. We first analyzed our data assuming that both $R_{0}$
and $\tau_{q}$ are $T$-independent. This is the usual assumption,
commonly made when the $T$-dependence of $R_{0}$ is small. As seen in Fig.\ 1, $R_{0}$ in our measurements has some temperature dependence. We then analyzed our data including the $T$-dependence of $R_{0}$ and assuming that the relative $T$-dependence of $\tau_{q}$ is equal to half the relative $T$-dependence of $R_{0}$ \cite{GokmenPRB2009,AdamovPRB2006}. We found that these two different methods yield very similar values for $m^{*}$ within our error bar range ($\sim$ 10$\%$), thus here we only report $m^{*}$ values deduced by assuming that both $R_{0}$ and $\tau_{q}$ are $T$-independent.

\section{Energy Band Calculations}
\label{sec:calc}

For comparison, we performed parameter-free calculations in the multiband envelope-function and self-consistent Hartree approximations for the quasi-2D hole system. \cite{win03} These calculations based on the $8 \times 8$ Kane Hamiltonian fully take into account the details of the 2D holes' band structure such as the nonparabolicity, anisotropy, and spin splitting of the energy dispersion $E_\pm (\bm{k}_\|)$. From the dispersion we obtain the density-of-states effective mass at the Fermi energy $E=E_F$
\begin{equation}
  \label{eq:mass}
  m^{*}_\pm (E_F) = \frac{\hbar^2}{2\pi\, m_e}
  \int \delta [E_F - E_\pm (\bm{k}_\|)] \:  d^2 k_\|
\end{equation}
which is the quantity we compare with the measured values of $m^{*}$. We evaluate Eq.\ (\ref{eq:mass}) by means of analytic quadratic Brillouin zone integration. \cite{win93} We note that the effective mass according to  Eq.\ (\ref{eq:mass}) equals the cyclotron effective mass in the limit $B \rightarrow 0$.

\section{Effective Mass in the Absence of Parallel Field}

\begin{figure}
\centering
\includegraphics[scale=0.85]{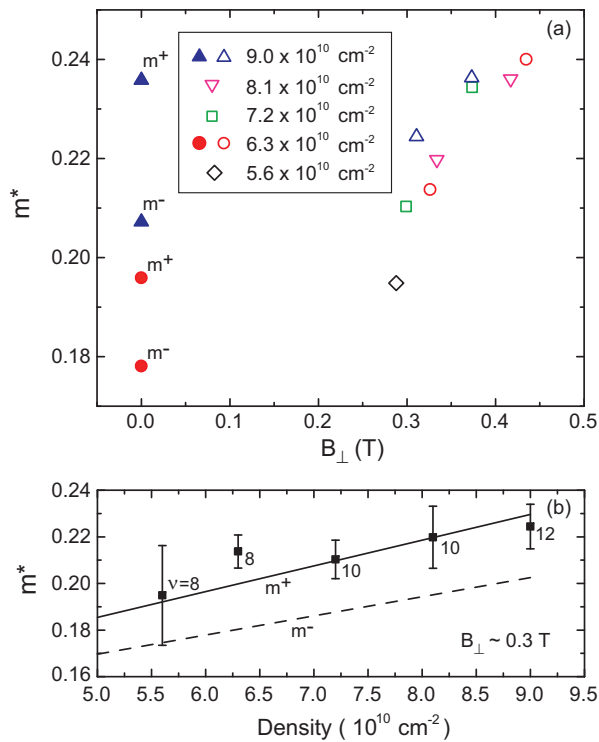}
\caption{(a) Effective mass ($m^{*}$) of dilute 2D holes at different densities. The experimentally measured $m^{*}$ are represented by open symbols and have an accuracy of $\pm$10$\%$. Closed symbols represent calculated values. (b) Experimentally measured $m^{*}$ at $B_{\bot} \sim 0.3 $ T as a function of density. The filling factor at which $m^{*}$ was measured is indicated for each data point. The solid and dashed lines represent the results of theoretical calculations (at $B_{\bot}=0$) for $m^{+}$ and $m^{-}$, respectively.}
\end{figure}

Figure 1 shows representative $R_{xx}$ vs. $B_{\bot}$ data for $p = 6.3 \times 10^{10}$ cm$^{-2}$. Because of the occupation of two spin subbands (as shown schematically in Fig.\ 1 right inset), and the order of magnitude smaller Zeeman energy compared to the cyclotron energy, the magneto-resistance traces in Fig.\ 1 exhibit stronger minima at even values of Landau level filling factors ($\nu$) compared to odd $\nu$. The minima at odd $\nu$ are indeed absent at low fields; this is the range where we analyze the temperature dependence of the SdH oscillations. The left inset in Fig.\ 1 shows the SdH oscillation amplitude fitted to the Dingle expression, \cite{footnote.Dingle.analysis} and assuming $T$-independent $\tau_{q}$ and $R_{0}$, the corresponding $m^{*}$ from this analysis is $m^{*} = 0.21$, in units of free electron mass $m_{e}$.

We performed similar measurements and analysis at various densities, obtained by biasing the front gate. The resulting $m^{*}$ are shown in Fig.\ 2. In the density range probed here $m^{*}$ lies between 0.19 and 0.24, with the higher values typically corresponding to higher densities (see Fig.\ 2 (b)). For each density, we also deduced $m^{*}$ at different filling factors, each corresponding to a specific $B_{\bot}$. As seen in Fig.\ 2 (a), the measured $m^{*}$ exhibits a slight increase with $B_{\bot}$. Similar trends have been previously reported \cite{EisensteinPRL1984, HabibPRB2004}. The origin is not fully clear but is likely related to the non-linear dependence of the 2D hole Landau levels on $B_{\bot}$ \cite{ColePRB1997, RachorPRB2009}. To asses the density dependence of $m^{*}$, we show a plot of $m^{*}$, measured at comparable values of $B_{\bot} \sim 0.3 $ T. The plot suggests a slight increase on $m^{*}$ with increasing density.

In Fig.\ 2 (a), we also include values for $m^{+}$ and $m^{-}$, effective masses calculated at $B=0$ for the two spin subbands for two sample densities, 6.3 and $9.0 \times 10^{10}$ cm$^{-2}$. The calculations are performed for the parameters of our sample, namely a 20-nm-wide GaAs quantum well, and assume that the charge distribution in the quantum well is symmetric \cite{ChargeDistributionAssymetry}. Note that the calculations predict slightly different densities $p^{+}$ and $p^{-}$, and corresponding effective masses $m^{+}$ and $m^{-}$, for the two spin subbands split by spin-orbit interaction. For $p = 6.30 \times 10^{10}$ cm$^{-2}$, e.g., the calculations indicate $p^{+} = 3.28 \times 10^{10}$ cm$^{-2}$ ($m^{+} = 0.196$) and $p^{-} = 3.02 \times 10^{10}$ cm$^{-2}$ ($m^{-} = 0.178$). In principle, the values of $p^\pm$ and $m^\pm$ can be measured at high densities by separating and analyzing the SdH oscillations associated with the spin-split subbands. Such analysis was indeed reported for 2DHSs with much higher densities and confined to a triangular confining potential so that the splitting between the $p^{+}$ and $p^{-}$ bands and the difference between $m^{+}$ and $m^{-}$ were much more pronounced \cite{HabibPRB2004}. In our low density sample with nearly equal $p^{+}$ and $p^{-}$, however, we are not able to experimentally resolve $p^{+}$ and $p^{-}$ or the two effective masses.

In Fig. 2(b) we have included two curves (solid and dashed) to represent the theoretical values of $m^{+}$ and $m^{-}$, calculated at $B_{\bot}=0$. Given the error bars of the experimental data, and the fact that $m^{*}$ shown in Fig. 2(b) were measured at $B_{\bot} \sim 0.3 $ T while $m^{+}$ and $m^{-}$ were calculated at $B_{\bot}=0$, there seems to be reasonable agreement between the measured and calculated effective masses, including the slight increase with density.

\begin{figure}
\centering
\includegraphics[scale=0.85]{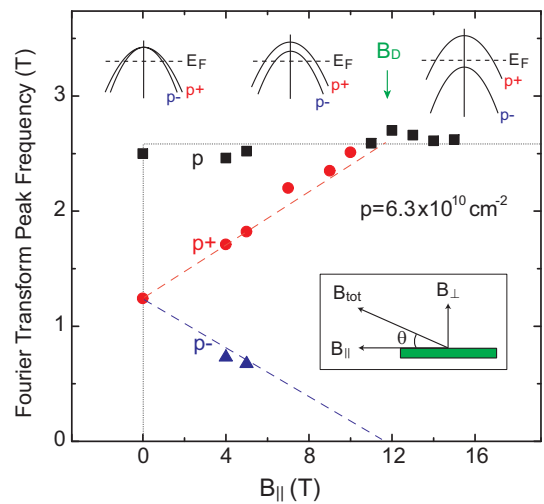}
\caption{Fourier transform peak frequencies of SdH
oscillations are plotted as a function of $B_{\|}$. The SdH
oscillations were measured by tilting the sample in a constant
magnetic field as shown in the bottom inset. Top insets show the evolution of the GaAs 2DHS band structure as a function of increasing $B_{\|}$. Square (black) symbols correspond to total density ($p$), and the circle (red) and triangle (blue) symbols to the densities of the $p^+$ and $p^-$ subbands. The dashed curves through the data points are guides to the eye. Top insets schematically show the evolution of the GaAs 2DHS band structure as a function of increasing $B_{\|}$.}
\end{figure}

\section{Effective Mass in a Strong Parallel Field}

A magnetic field applied parallel to the 2DHS leads to an enhancement of the Zeeman splitting between the energies of the spin subbands and, for a sufficiently large $B_{\|}$, the holes are all transferred to the $p^{+}$ band (see Fig.\ 3 top insets) \cite{TutucPRL2001}. Following the experimental procedure of Ref.~~\onlinecite{TutucPRL2001}, we first applied a strong magnetic field parallel to the $[\bar{2}33]$ direction of our 2DHS. We then slowly tilted the sample to introduce a small perpendicular magnetic field component ($B_{\bot}$) and measured the resulting SdH oscillations as the sample was being tilted (see Fig.\ 3 lower inset). An example of the resulting SdH oscillations is shown in Fig.\ 4 (a). Note that in this procedure, the total applied magnetic field ($B_\mathrm{tot})$ is constant and, if the angle $\theta$ between $B_\mathrm{tot}$ and the sample plane is small ($<$ 5$^\circ$), $B_{\|}$ is nearly equal to $B_\mathrm{tot}$ and is therefore essentially a constant over the whole tilting range \cite{TutucPRL2001}.

For sufficiently large values of $B_{\|}$, the Fourier transform of the SdH oscillations resulting from $B_{\bot}$ exhibits two peaks corresponding to the populations of the $p^+$ and $p^-$ subbands, as well as a third peak which is a measure of the total density in the 2DHS \cite{TutucPRL2001}. Figure 3 shows the evolution of these three peaks as a function of $B_{\|}$ in our sample at a total density of $p = 6.3 \times 10^{10}$ cm$^{-2}$. Note that for $B_{\|} <$  4 T, the splitting of the two spin subbands is small and we could not resolve them in the Fourier transform spectrum. As $B_{\|}$ changes from 4 to 15 T, $p^-$ decreases while $p^+$ grows and eventually saturates at a value corresponding to the total density. The field $B_D$ (about 11.5 T for the data of Fig.\ 3, as marked by a vertical arrow) signals the complete depletion of the $p^-$ subband and full spin subband polarization of the 2DHS \cite{TutucPRL2001}.

\begin{figure}
\centering
\includegraphics[scale=0.85]{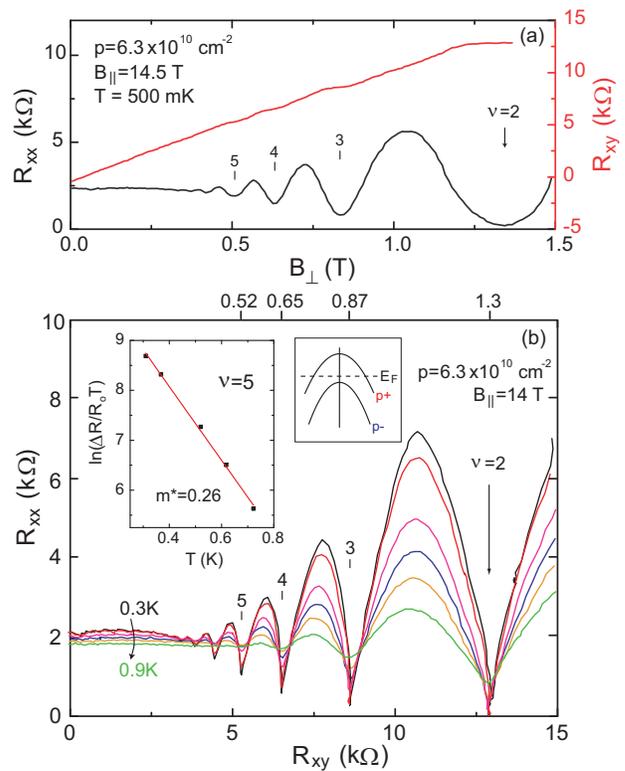}
\caption{(a) Longitudinal ($R_{xx}$) and Hall ($R_{xy}$) resistance traces taken at a high parallel magnetic filed of 14.5 T and plotted versus the perpendicular component of the magnetic field. (b) $R_{xx}$ plotted versus $R_{xy}$ at a density of $6.3 \times 10^{10}$ cm$^{-2}$ and parallel magnetic field of 14 T. The top axis indicates the values of $B_{\bot}$ for the filling factors $\nu = 2, 3, 4,$ and 5. The traces are taken at various temperatures from 0.3 to 0.9 K. Left inset shows the fit of the SdH oscillation amplitude at $\nu = 5$ to the Dingle expression. The right inset schematically shows the fully-polarized spin subbands for the 2DHS.}
\end{figure}

In order to determine $m^{*}$ in our 2DHS when it is fully spin-subband polarized, we applied a magnetic field larger than $B_D$ and then tilted the sample slightly to induce SdH oscillations (see Fig.\ 4 (a)) and measured the temperature dependence of these oscillations. An example of such set of data is presented in Fig.\ 4 (b), where we show traces of $R_{xx}$ versus $R_{xy}$ for $p = 6.3 \times 10^{10}$ cm$^{-2}$ at $B_{\|} = 14$ T and at different temperatures. Since the 2D holes at this $B_{\|}$ occupy only one spin subband, namely the $p^+$ subband (see Fig.\ 4 (b) right inset), the SdH oscillations exhibit a simple pattern; notably, minima are clearly visible for both odd and even filling factors. To determine $m^{*}$, similar to the $B_{\|} = 0$ measurements, we fit the amplitude of the SdH oscillations for a given filling factor to the Dingle expression assuming $T$-independent $\tau_{q}$ and $R_{0}$. An example is shown in Fig.\ 4 (b) (left inset) for the SdH oscillation at $\nu = 5$, from which we deduce $m^{*} = 0.26$. In Fig.\ 5, we show the measured $m^{*}$ at different values of $B_{\|} > B_D$ for $p = 6.3 \times 10^{10}$ cm$^{-2}$. At each value of $B_{\|}$, $m^{*}$ was determined at $\nu = 3, 4$, and 5, as indicated in Fig.\ 5 by different colors. In Fig.\ 5 we also show similarly determined $m^{*}$ at a density of $3.7 \times 10^{10}$ cm$^{-2}$.

Data of Fig.\ 5 indicate that the measured $m^{*}$ does not show a systematic dependence on $B_{\|}$ within our experimental accuracy. Interestingly, however, we find that $m^{*}$ appears to be somewhat larger at lower fillings, i.e., it slightly increases with increasing $B_{\bot}$. This behavior is very similar to what is seen in the absence of $B_{\|}$ (see Fig.\ 2(a)). To highlight the dependence of $m^{*}$ on $B_{\bot}$, in Fig.\ 6 we show a summary of $m^{*}$ measured at different densities for $B_{\|} = 0$ (open symbols) \cite{lowdensity}, and at very large values of $B_{\|}$ ($> B_D$) so that the 2DHS is fully spin-subband polarized (closed symbols). In the fully polarized case, at each value of $B_{\bot}$, in Fig.\ 6 we show the average value of $m^{*}$ in the range of $B_{\|}$ where $m^{*}$ was measured; the error bars in Fig.\ 6 include the variation of $m^{*}$ as a function of $B_{\|}$.

\begin{figure}
\centering
\includegraphics[scale=0.85]{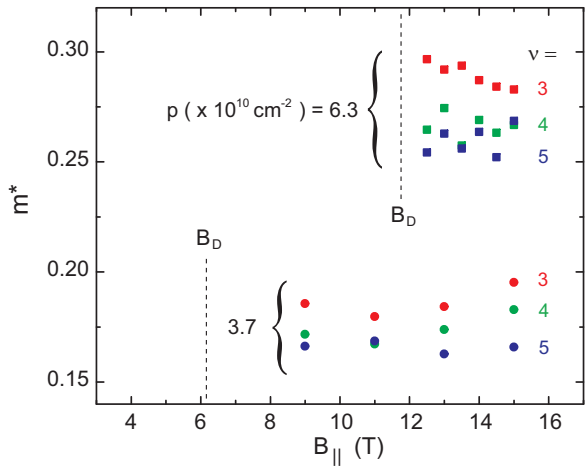}
\caption{ (Color online) Effective mass ($m^{*}$) plotted as a function of the parallel magnetic field ($B_{\|}$) when the 2DHS is fully spin-subband polarized. Different colors represent $m^{*}$ measured at different filling factors; the experiment accuracy for $m^*$ is about $\pm 5 \%$. Data are shown for two different densities $p = 6.3 \times 10^{10}$ cm$^{-2}$ and $3.7 \times 10^{10}$ cm$^{-2}$. Note that, for each density, $B_{\|}$ at which $m^{*}$ was measured is larger than the corresponding $B_D$ for that density. }
\end{figure}

\begin{figure}
\centering
\includegraphics[scale=0.85]{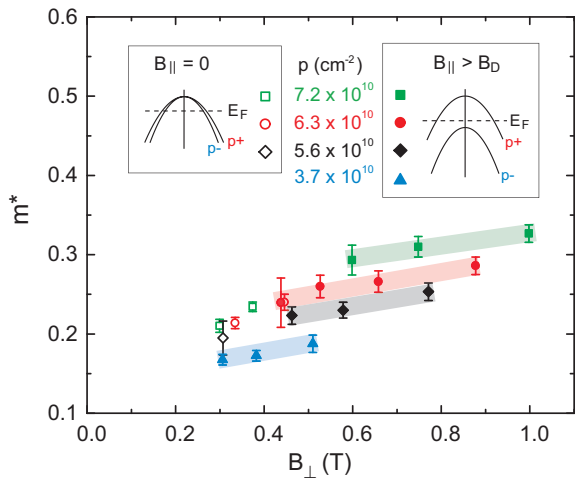}
\caption{Effective mass ($m^{*}$) measured as a function of perpendicular magnetic field ($B_{\bot}$). Open symbols correspond to the case of $B_{\|} = 0$ and the closed symbols to the case where $B_{\|} > B_D$ so that the 2DHS is fully- spin-subband polarized. Different colors represent $m^{*}$ measured at different densities. Insets schematically show the spin subbands of the 2DHS. Color bands are guides to the eye, and suggest that $m{*}$ measured at $B_{\|} > B_D$ also increases with increasing density.}
\end{figure}

\section{Discussion of Effective Mass Data}

We first consider data of Fig.\ 2 which were taken at $B_{\|} = 0$. From these data we draw two main conclusions: (1) The values of $m^{*}$ determined experimentally are, within the accuracy of the measurements, in reasonable agreement with the calculated $m^{*}$. (2) The measured $m^{*}$ appear to be slightly larger for higher densities, again, this is consistent with the results of calculations (Fig. 2(b)).

Given that the calculated $m^{*}$ are deduced from band calculations that do not include interaction effects \cite{CR.comparison}, it appears then that there is no appreciable enhancement of $m^{*}$. This is surprising. Previous studies on 2D electron systems confined to Si-MOSFETs \cite{ShashkinPRB02}, to GaAs \cite{TanPRL2005,TanPRB2006}, or to narrow (width $<$ 5 nm) AlAs quantum wells \cite{GokmenPRB2010} have reported an enhancement of $m^{*}$ by about 50$\%$ over the band value for a comparable value of $r_{s}$ ($\sim $ 6). The reason for the lack of $m^{*}$ enhancement in 2D holes is not obvious but is possibly related to the holes' band structure and effective spin $j =  3/2$, as discussed in Ref.~~\onlinecite{WinklerPRB2005}. We remark, however, that negligible $m^{*}$ enhancement has also been seen in another 2D system, namely in electrons confined to wide (width $> 5$ nm) AlAs quantum wells where a conduction-band valley with an anisotropic in-plane contour is occupied [see Figs.\ 1 and 2(e) of Ref.~~\onlinecite{GokmenPRB2010}]. Another potentially important factor is the role of disorder. According to theoretical calculations, a larger $m^{*}$ is expected for a more disordered 2D carrier system \cite{AsgariSSC2004} but there have been no systematic experimental studies assessing the influence of disorder.

Next, we consider data taken at large $B_{\|}$ where the 2D holes occupy only the $p^+$ spin subband (Figs.\ 5 and 6 data). The data of Fig.\ 6 indicate that applying a strong $B_{\|}$ does not seem to affect $m^{*}$. There is some enhancement of $m^{*}$ at large $B_{\|}$ (closed symbols) with respect to the $B_{\|} = 0 $ values (open symbols), but this enhancement appears to be correlated with the increase of $m^{*}$ with the \textit{perpendicular} component of field rather than a dependence on $B_{\|}$. This conjecture is corroborated by the data of Fig.\ 5 which suggest no significant or systematic dependence of $m^{*}$ on $B_{\|}$.  The lack of a fairly large enhancement of $m^{*}$ at high values of $B_{\|}$ is also puzzling. Self-consistent calculations of $m^*$ for $B_\| > 0$ based on the approach described in Section~\ref{sec:calc} indicate a significant enhancement of $m^*$ at $B_D$ by about a factor of three. This increase reflects the more complicated nonparabolic dynamics of holes. Indeed, this increase is yet significantly larger than the increase of $m^*$ one would expect for an electron system with nominally similar values of $m^*$ and $g^*$ due to the coupling of $B_\|$ to the holes' orbital motion \cite{TutucPRB2003}. In the latter case we would expect a roughly 50$\%$ enhancement of $m^{*}$ (compared to its $B_{\|} = 0$ value). On the other hand, according to recent measurements \cite{PadmanabhanPRL2008, GokmenPRL2008, GokmenPRB2009, GokmenPRB2010} whose results were supported by subsequent calculations \cite{AsgariPRB2009, DrummondPRB2009}, $m^{*}$ is $suppressed$ in a 2D electron system which is fully spin polarized. The latter results were obtained for more narrow electron systems so that the coupling of $B_\|$ to the electrons' orbital motion was a small effect.

\begin{figure}
\centering
\includegraphics[scale=0.85]{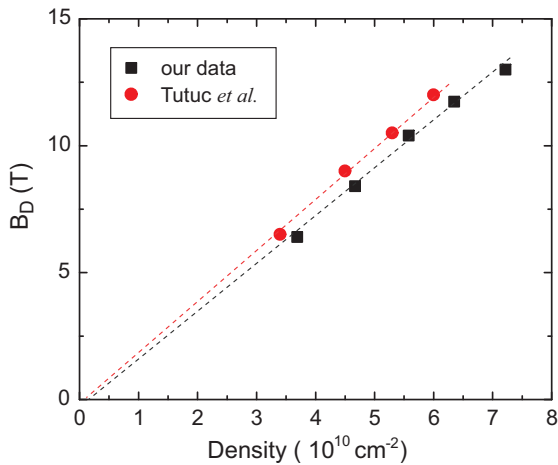}
\caption{Parallel magnetic field required to fully spin-subband polarize the 2DHS ($B_{D}$) versus density. The squares are data from our measurements. The circles represent data from Tutuc $et$ $al$ \cite{TutucPRL2001}. The two dashed lines are least-squares fits through each set of data points.}
\end{figure}

\section{Spin Susceptibility of Dilute 2D Holes}

From our parallel field experiments we can also deduce a value for the spin susceptibility ${\chi}^*$ which, in a 2D carrier system, is proportional to the product $g^{*} m^{*}$. At the magnetic field $B_D$ where the minority spin subband is depopulated, the Zeeman energy is equal to the Fermi energy of the 2D system. If $m^{*}$ and $g^{*}$ are independent of $B_{\|}$ then the depopulation of the minority spin subband is linear with $B_{\|}$, and the equality of the Zeeman and Fermi energies leads to the expression: $B_D = (h^2/2{\pi}{\mu}_B) \cdot (p/g^{*}m^{*})$, where ${\mu}_B$ is the Bohr Magneton. The value of $B_D$ therefore provides a direct measure of $g^{*}m^{*}$. In Fig.\ 3, as well as in Fig.\ 2 of Ref.~~\onlinecite{TutucPRL2001}, we observe that $p^+$ indeed does depend nearly linearly on $B_{\|}$ (for $B_{\|}$ applied along the $[\bar{2}33]$ direction), suggesting that $g^{*}m^{*}$ does not depend on $B_{\|}$ \cite{B.parallel}. Moreover, Fig.\ 7 shows that the measured field $B_D$ has approximately a linear dependence on density, and that a line fitted through the data points nearly passes through the origin. Using the above expression for $B_D$, the slope of this line yields $g^{*}m^{*}$ $\cong$ 0.19. Considering the band values $m^{*}$ $\cong$ 0.2 and $g^{*}$ $\cong$ 0.65 \cite{WinklerPRL2000}, this implies an enhancement of $g^{*}m^{*}$ by a factor of about 1.5. Such an enhancement is about a factor of two smaller than the $g^{*}m^{*}$ enhancement reported for 2D electrons in Si-MOSFETs \cite{ShashkinPRB02}, GaAs \cite{ZhuPRL2003}, or AlAs \cite{GokmenPRB2010} at comparable values of $r_s$. The reason for this absence of enhancement might again be the holes' band structure and large effective spin \cite{WinklerPRB2005}.

\section{Summary and Conclusions}

Our effective mass measurements illustrate that $m^{*}$ in dilute 2D holes confined to GaAs quantum wells has a value close to 0.2, and increases slightly with increasing density. Both the magnitude of $m^{*}$ and its density dependence are in agreement with the results of energy band calculations for our 2DHS. We also apply a strong parallel magnetic field to depopulate the minority spin subband and measure $m^{*}$ for the majority spin subband.  We find that $m^{*}$ is not influenced substantially by the large parallel field. Finally, we deduce the spin susceptibility of the 2DHS from the depopulation field, and conclude that the susceptibility is enhanced by about 50$\%$ relative to the value expected from the band calculations. The lack of significant enhancements of $m^{*}$ and the susceptibility possibly originates from the holes' band structure and $j = 3/2$ spin \cite{WinklerPRB2005}.

\begin{acknowledgments}
  We acknowledge support through the Department of Energy (Grant
  DEFG02-00-ER45841) for sample fabrication, and the National
  Science Foundation (Grants MRSEC DMR-0819860, ECCS-1001719 and
  0829872) for characterization and measurements. Work at Argonne
  was supported by DOE BES under Contract No.\ DE-AC02-06CH11357.
\end{acknowledgments}

%\break


\begin{references}
\bibitem{KohnPR1964}
W.\ Kohn, Phys.\ Rev.\ {\bf 123}, 1242 (1961). However, it is important
to note that Kohn's theorem is applicable only if the particles have a
parabolic energy vs. wave-vector dispersion, which is not the case for
holes.
\bibitem{Footnote1}
2DHSs grown on GaAs (311)A substrates exhibit a mobility anisotropy stemming from an anisotropic surface morphology. The interfaces between GaAs and AlGaAs are corrugated, with ridges along the $[\bar{2}33]$ directions, thus reducing the mobility for current parallel to $[01\bar{1}]$.
\bibitem{footnote.Dingle.analysis}
Because of the moderate mobility and the very low densities of our sample, we have to perform our $m^{*}$ determination at relatively small filling factors $\nu<13$. We would like to emphasize that the measured $m{*}$ we report throughout this paper were determined from data which fit the Dingle expression very well in the entire temperature range where the fit was done.
\bibitem{ChargeDistributionAssymetry}
Since we change the density via front gate bias, we expect that for some densities the charge distribution is somewhat asymmetric. Our calculations performed for reasonable values of such asymmetry indicate $m^{*}$ values which differ from the symmetric case by only a few percent.
\bibitem{lowdensity}
For the lowest density shown in Fig.\ 6, the SdH oscillations are too weak to deduce $m^*$ at $B_{\|} = 0$.
\bibitem{CR.comparison}
Both the values, and the slight enhancement with increasing density, of $m^{*}$ that we obtain from the band calculations are also consistent with the results of cyclotron resonance measurements (see, e.g., Refs.~~\onlinecite{ColePRB1997} and ~~\onlinecite{RachorPRB2009}).
\bibitem{B.parallel}
Note that the conjecture that $m^{*}$ is nearly independent of $B_{\|}$ is consistent with our $m^{*}$ data shown in Fig.\ 5.
\end{references}
\end{document}